\documentclass[a4paper,11pt]{article}
\usepackage{pos}
\usepackage{deluxetable}
\usepackage{caption}
\usepackage{subcaption}
\usepackage{soul}

\graphicspath{{./}{figures/}}

\title{Energy Balance at Interplanetary Shocks: In-situ Measurement of the Fraction in Supra-thermal and Energetic Particles with ACE and Wind}
 \ShortTitle{Energy Balance at Interplanetary Shocks}

\author*[a]{Liam David}
\affiliation[a]{Lunar \& Planetary Laboratory, University of Arizona, Tucson, AZ 85721, USA}

\author[a,b]{Federico Fraschetti}
\affiliation[b]{Center for Astrophysics | Harvard \& Smithsonian, Cambridge, MA, 02138, USA}

\author[a]{Joe Giacalone}

\author[c,d]{Robert Wimmer-Schweingruber}
\affiliation[c]{Institute of Experimental and Applied Physics, Kiel University, Kiel, Germany}
\affiliation[d]{National Space Science Center, Chinese Academy of Sciences, Beijing, China}

\author[c]{Lars Berger}

\author[e]{David Lario}
\affiliation[e]{Heliophysics Science Division, NASA Goddard Space Flight Center, Greenbelt, MD, USA}

\emailAdd{liamdavid@email.arizona.edu}

\abstract{The acceleration of charged particles by interplanetary shocks can drain a non-negligible fraction of the upstream ram pressure. For a sample of shocks observed in-situ at 1 AU by the ACE and Wind spacecraft, time-series of the non-Maxwellian components (supra-thermal and higher-energy) of the ion and electron energy spectra were acquired for each event. These were averaged for one hour before and after the time of the shock passage to determine their partial pressure. Using the MHD Rankine-Hugoniot jump conditions, we find that the fraction of the total upstream energy flux transferred to non-Maxwellian downstream particles is typically about 2-16\%. Notably, our sample shows that neither the fast magnetosonic Mach number nor the angle between the shock normal and average upstream magnetic field are correlated with non-Maxwellian particle pressure.}

\FullConference{37$^{\rm{th}}$ International Cosmic Ray Conference (ICRC 2021)\\
		July 12th -- 23rd, 2021\\
		Online -- Berlin, Germany}
\begin{document}
\maketitle

\section{Introduction}
Shocks in astrophysical plasmas have been long known to be efficient producers of high-energy charged particles. The association between interplanetary shocks in the solar wind at 1 AU and intensity enhancements of $>50$ keV/nuc ions has been reported for decades \citep{Scholer.etal:1983, Reames:99,Desai.Giacalone:16}, although few studies have focused on quantifying the fraction of the upstream ram pressure transferred into energetic particles. An example is the crossing of the solar wind termination shock by Voyager 2: a measured plasma temperature roughly $10$ times smaller than predicted was interpreted as transfer of a large fraction of the solar wind energy into the pick-up ions \cite{Richardson.etal:2008}.

As for the inner solar wind, Mewaldt et al. \cite{Mewaldt.etal:08} estimated that in large (kinetic energy $>$$10^{31}$ erg) coronal mass ejections (CMEs) the fraction of the bulk kinetic energy transferred into accelerated particles can attain values as high as $10 - 20 \%$. On larger scales, shocks produced by supernova explosions also generate significant energetic particle fluxes: Slane et al. \cite{Slane.etal:14} inferred that $\sim$$16\%$ of the bulk kinetic energy is converted into accelerated particles in the {\it Tycho}'s supernova remnant shock by using a hydrodynamic model of the broadband spectrum (from radio to multi-TeV radiation).

In this study, we examine proton and electron acceleration at interplanetary shocks at 1 AU, as measured {\it in-situ} by the Advanced Composition Explorer (ACE) and the Wind spacecraft. By combining time-series of the suprathermal components and the high-energy tails of the proton and electron distribution functions with the background plasma parameters, we determine the energy fraction swept up by a shock that is transferred into the non-Maxwellian populations. 

\section{Shock Data}
The Center for Astrophysics | Harvard \& Smithsonian (CfA) shocks database catalogues interplanetary shocks observed by the ACE and Wind spacecraft, and provides shock parameters including the time, speed, and local normal direction, as well as plasma parameters such as proton temperatures, proton densities, and magnetic fields both upstream and downstream; we refer below to values determined with the RH08 method \cite{Koval.Szabo:08}. For dynamically varying quantities such as the magnetic field and density, the database uses adaptive averaging within roughly $\pm20$ minutes of the shock time.

We selected eight fast-forward shocks observed by ACE and ten by Wind between 1997 and 2013. These are listed in Table 1. Since errors on the shock speed and normal direction have the greatest effect on our final uncertainties, they were used as selection criteria. In addition, we restricted our analysis to events with a density compression greater than $2$ and fast magnetosonic Mach number $M_{ms}>1.25$. In order to examine the dependence of the partial pressure on $M_{ms}$ and $\theta_{Bn}$, i.e., the angle between the upstream magnetic field and the shock normal, we selected events spanning the broadest possible range of values: $M_{ms}$ from $1.49$ to $6.21$, as the shock speed uncertainty increases with $M_{ms}$, and $\theta_{Bn}$ from $19.2^\circ$ to $89.8^\circ$. In addition, the energetic particle intensity profiles, which are discussed next, were required to have an exponential-like pre-shock rise, with small-amplitude fluctuations, followed by a flat top in the downstream plasma.

\begin{deluxetable}{lll}
\tablecaption{Interplanetary shocks used in this study from ACE (A) and Wind (W). Values and errors on $\theta_{Bn}$ are taken from the CfA database. Values of  $M_{ms}$ are calculated from the shock speed and fast plasma speed in the CfA database and errors linearly propagated.\label{chartable}}
\tablewidth{0pt}
\tabletypesize{\footnotesize}
\tablehead{
    \colhead{(Spacecraft Year/Day/UT) [DOY]} & \colhead{$M_{ms}$} & \colhead{$\theta_{Bn}$}
}
\startdata
A 2001/298/8:2:1 [298.3347] & 5.6 $\pm$ 0.7  & 40.0 $\pm$ 7.9 \\ 
A 2005/21/16:47:26 [21.6996] & 5.3 $\pm$ 0.6  & 69.8 $\pm$ 17.8 \\ 
A 1999/49/2:8:50 [49.0895] & 3.6 $\pm$ 0.2  & 50.8 $\pm$ 7.4 \\ 
A 1999/346/15:14:36 [346.6352] & 1.6 $\pm$ 0.1  & 68.3 $\pm$ 10.0 \\ 
A 2001/23/10:6:20 [23.4211] & 2.9 $\pm$ 0.4  & 19.2 $\pm$ 8.3 \\ 
A 2001/86/17:15:15 [86.7189] & 1.5 $\pm$ 0.2  & 74.4 $\pm$ 3.6 \\ 
A 2001/118/4:31:58 [118.1889] & 4.5 $\pm$ 0.7  & 89.7 $\pm$ 4.1 \\ 
A 2001/229/10:16:2 [229.4278] & 2.7 $\pm$ 0.3  & 79.8 $\pm$ 5.0 \\ 
W 2003/308/6:46:4 [308.2820] & 2.7 $\pm$ 0.4  & 63.9 $\pm$ 11.7 \\ 
W 1997/326/9:12:52 [326.3839] & 2.3 $\pm$ 0.1  & 83.3 $\pm$ 3.6 \\ 
W 2013/275.1/1:15:49 [275.0527] & 4.2 $\pm$ 0.2  & 28.6 $\pm$ 8.3 \\ 
W 2004/312/17:59:5 [312.7494] & 2.0 $\pm$ 0.2  & 60.9 $\pm$ 3.4 \\ 
W 1998/275.3/7:6:4 [275.2959] & 2.5 $\pm$ 0.9  & 26.2 $\pm$ 9.9 \\ 
W 2013/76/5:21:28 [76.2232] & 5.9 $\pm$ 0.4  & 35.1 $\pm$ 5.8 \\ 
W 2002/77/13:14:4 [77.5514] & 5.7 $\pm$ 0.9  & 44.2 $\pm$ 20.3 \\ 
W 2004/208/22:25:23 [208.9343] & 5.4 $\pm$ 0.7  & 57.8 $\pm$ 7.0 \\ 
W 2005/252/13:33:1 [252.5646] & 6.2 $\pm$ 0.5  & 89.8 $\pm$ 8.6 \\  
\enddata
\end{deluxetable}

\section{Energetic Particle Data}

Proton data for the ACE shocks were retrieved from the Electron, Proton, and Alpha Monitor (EPAM) and the Solar Wind Ion Composition Spectrometer (SWICS) instruments through the CalTech Level 2 Database.
EPAM contains a Low Energy Magnetic Spectrometer, designated as LEMS120, that points 120$^{\circ}$ off the spacecraft spin axis, i.e., out of the sun line-of-sight. It records ion intensities between $47.0$ keV and $4.80$ MeV in $8$ bins, integrating fluxes in parallel for 12 seconds to constitute one sample. We assume the measured fluxes are dominated by the more abundant protons. SWICS is a linear time-of-flight (TOF) mass spectrometer with electrostatic deflection, allowing it to measure ion mass, charge, and energy. A full energy spectrum covering $657$ eV - $86.6$ keV in $58$ bins is collected every 12 minutes; each bin is integrated serially for 12 seconds. Here the background thermal solar wind, suprathermal protons, and heavier ions can be disentangled, and we have considered just proton observations.

Electron data for the ACE shocks were collected by the Deflected Electrons (DE30) instrument, which points 30$^\circ$ off the spin axis, through the CalTech Level 2 Database. Four energy bins span 38.0 keV - 315 keV and the time resolution is 12 seconds. A magnetic deflector separates these electrons from the parent LEMS30 instrument, which are then measured by a totally depleted surface barrier silicon detector.

Omnidirectional proton fluxes for the Wind events were obtained by the Three-Dimensional Plasma and Energetic Particle Investigation instrument (3DP) through the NASA CDAWeb database. The relevant detector is a solid-state telescope with a $36^\circ\times20^\circ$ field of view $126^\circ$ off the spin axis \cite{3dp}. Nine ion energy bins span $67.3$ keV - $6.75$ MeV and are integrated in parallel for 12-second samples.

We have also considered proton fluxes in the range 400 eV - 19.1 keV from the Wind/3DP PESA-Hi (PH) data product. Since diffusive shock acceleration operates only for particles with speeds  $v\gtrsim3V_{sh}$, where $V_{sh}$ is the shock speed, we had to remove all but one or two of the highest energy bins. The uncertainties in the energy bins, as well as large errors in the energy flux resulting from the data coarseness, led us to discard this energy range. %\ld{perhaps reword paragraph}

Wind 3DP also contains two electron data products collected by the high-range Electron Electrostatic Analyzer (EH) and Solid State Foil Telescope (SF) instruments: EHSP spans 136 eV - 27.6 keV in 15 energy bins and SFSP covers 27.0 keV - 517 keV in 7 bins. Data with a $\sim$24s resolution were obtained from the NASA CDAWeb database. 

\section{Energetic Particle Pressure}
The densities and pressures for the energetic particle species $s$ were found by integrating the flux distributions, assumed to be isotropic, over energy $E_s$ and momentum $p_s$ spaces:
\begin{equation}
    \rho_s=4\pi m_s\int_{p_{min}} ^{p_{max}} Fdp_s^{(sh)}
\end{equation}
\begin{equation}
    P_s=\frac{4}{3}\pi m_s\int_{E_{min}} ^{E_{max}} \gamma v_s^{(sh)}F dE_s^{(sh)} \, .
\end{equation}
Here $m_s$ is the mass of the particle of species $s$, i.e., proton or electron, $F$ is the flux in $\rm{cm^{-2}s^{-1}sr^{-1}MeV^{-1}}$, $v_p$ is the speed, $\gamma$ is the particle relativistic Lorentz factor and the superscript $(sh)$ denotes that particle momenta and energies have been relativistically transformed from the spacecraft frame to the shock frame in order to apply RH jump conditions. The values $p_{min}$ and $E_{min}$ are chosen so that the particle speed exceeds $\sim 3$ times the shock speed and can thus undergo diffusive shock acceleration.

The downstream particle pressures were found by averaging each energy bin for one hour, beginning one minute after the shock time. This choice matches the start but not the duration of 20 minute analysis time-window for the CfA plasma quantities; however, we note that the time resolution of the SWICS sampling throughout the energy band (12 minutes) forces us to use a one hour-average for energetic particles: five SWICS samples were deemed necessary to reduce the effects of fluctuations in the data, thereby also requiring 300-sample averages for EPAM, DE30, and 3DP, and 150-sample averages for EHSP and SFSP, to maintain time consistency across instruments. 

\section{MHD Jump Conditions}
The Rankine-Hugoniot (RH) jump conditions express the conservation of mass, momentum, energy, and magnetic flux of a magnetized fluid heated by a shock. The upstream and downstream plasma velocities are transformed from the spacecraft frame, as provided by the CfA database, into the shock frame. The conservation of energy flux across the shock discontinuity is then given by the jump condition
\begin{equation}
    \left[v_k\left(\underbrace{\rho_gh_g}_\text{thermal} +\underbrace{\frac{1}{2}\rho_gv^2}_\text{kinetic}+\underbrace{\sum_s \rho_s \,h_s }_\text{downstream energetic particles}\right)+\frac{1}{4\pi}\left(v_kB^2-B_k(\textbf{v}\cdot\textbf{B})\right)\right]=0 \, .
    \label{eq:en_cons}
\end{equation}
Here $ h_{g,s}=\frac{\Gamma}{\Gamma-1}\frac{P_{g,s}}{\rho_{g,s}}$ are specific enthalpies for the gas (only protons that dominate the plasma temperature) and the species $s$, $\textbf{v}$ is the plasma velocity, and subscript $k$ denotes projections of that quantity onto the shock normal. Since we assume the solar wind and energetic particle fluxes contain only non-relativistic protons and electrons, an adiabatic index $\Gamma=5/3$ was used. A more compressible, i.e., trans-relativistic, energized electron gas does not affect the result significantly. The brackets $[\cdot]$ indicate the difference of $(\cdot)$ between upstream and downstream, except for the energized particle partial pressures, calculated downstream only.

\begin{figure}
     \centering
     \begin{subfigure}[b]{0.49\textwidth}
         \centering
         \includegraphics[width=\textwidth]{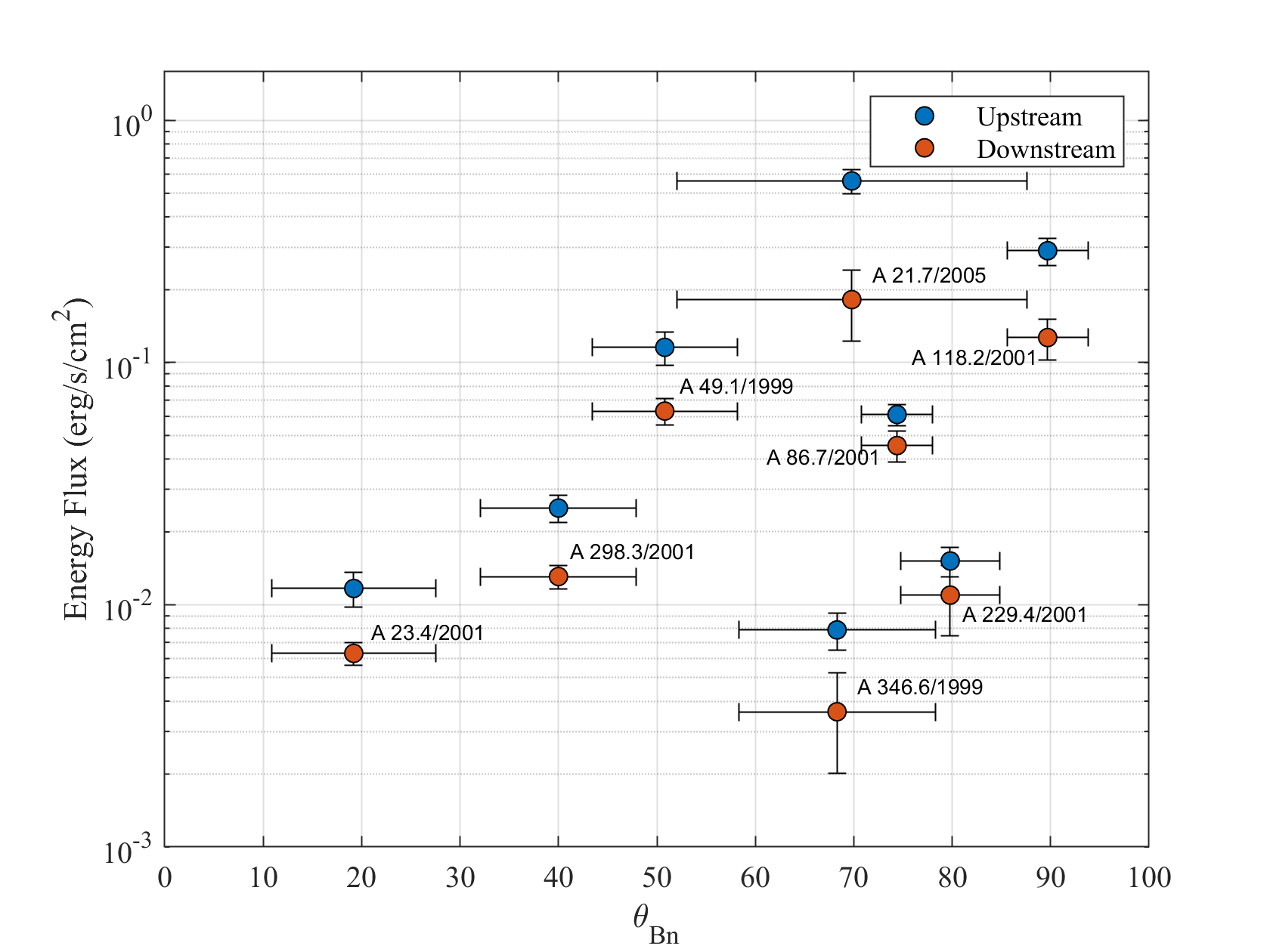}
         \caption{ACE shocks.}
     \end{subfigure}
     \begin{subfigure}[b]{0.49\textwidth}
         \centering
         \includegraphics[width=\textwidth]{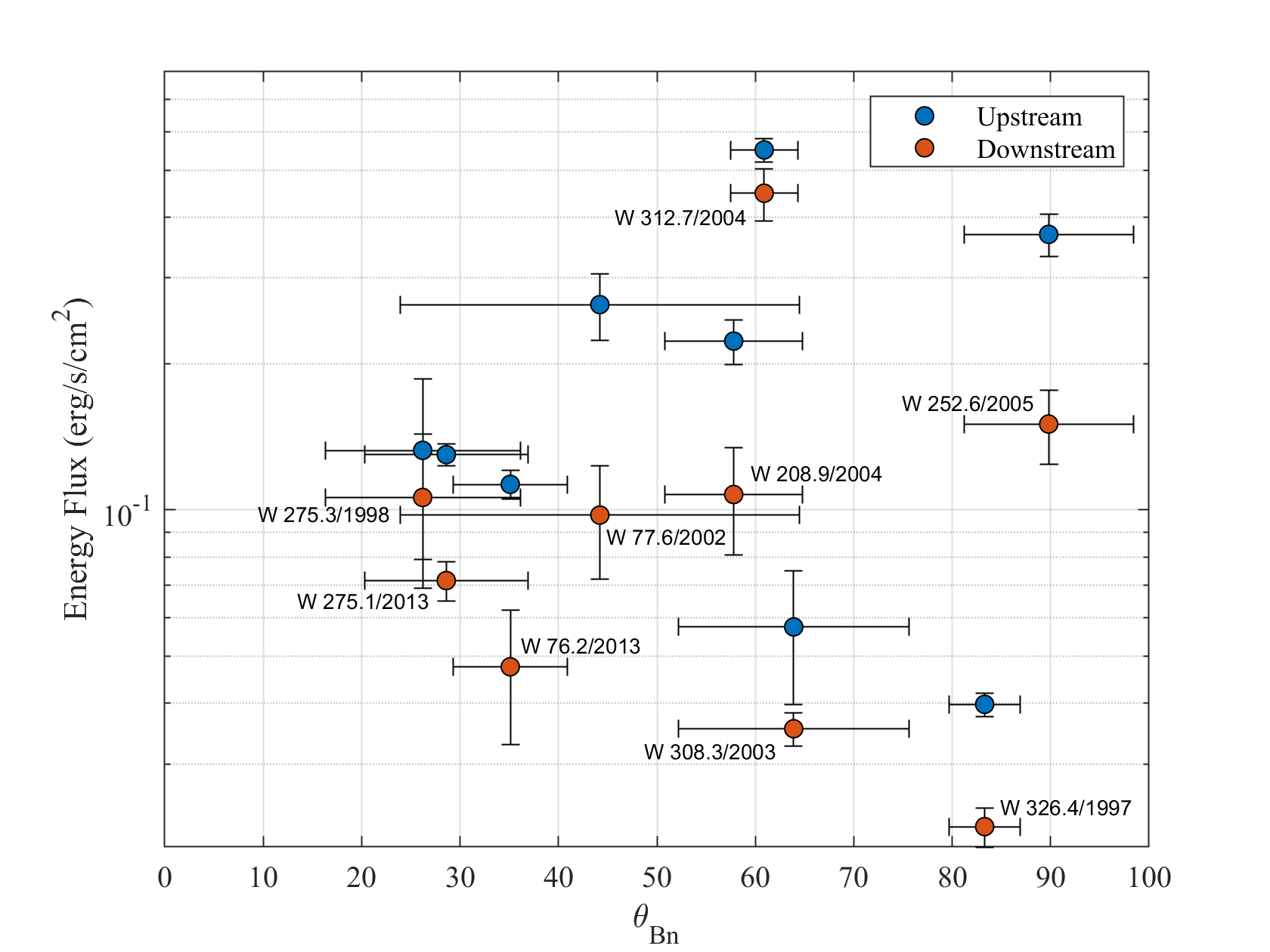}
         \caption{WIND shocks.}
     \end{subfigure}
        \caption{Upstream (blue) and downstream (red) total energy fluxes, including downstream energetic particle pressures (see Eq.\ref{eq:en_cons}), for each shock ((a) for ACE and (b) for Wind) as a function of $\theta_{Bn}$ with $1\sigma$ error bars. The downstream points are labeled with the shock DOY.\label{fig:jumpcondition}}
\end{figure}

The overall energy conservation is summarized in Figure \ref{fig:jumpcondition}. Only 1 out of the 8 ACE shocks and 1 out of the 9 Wind shocks show a matching of upstream and downstream energy fluxes within $1\sigma$ errors. Of note is that all upstream values are greater than those downstream, implying missing energy components, likely in the form of heavy ions.

\begin{figure}
     \centering
     \begin{subfigure}[b]{0.45\textwidth}
         \centering
         \includegraphics[width=\textwidth]{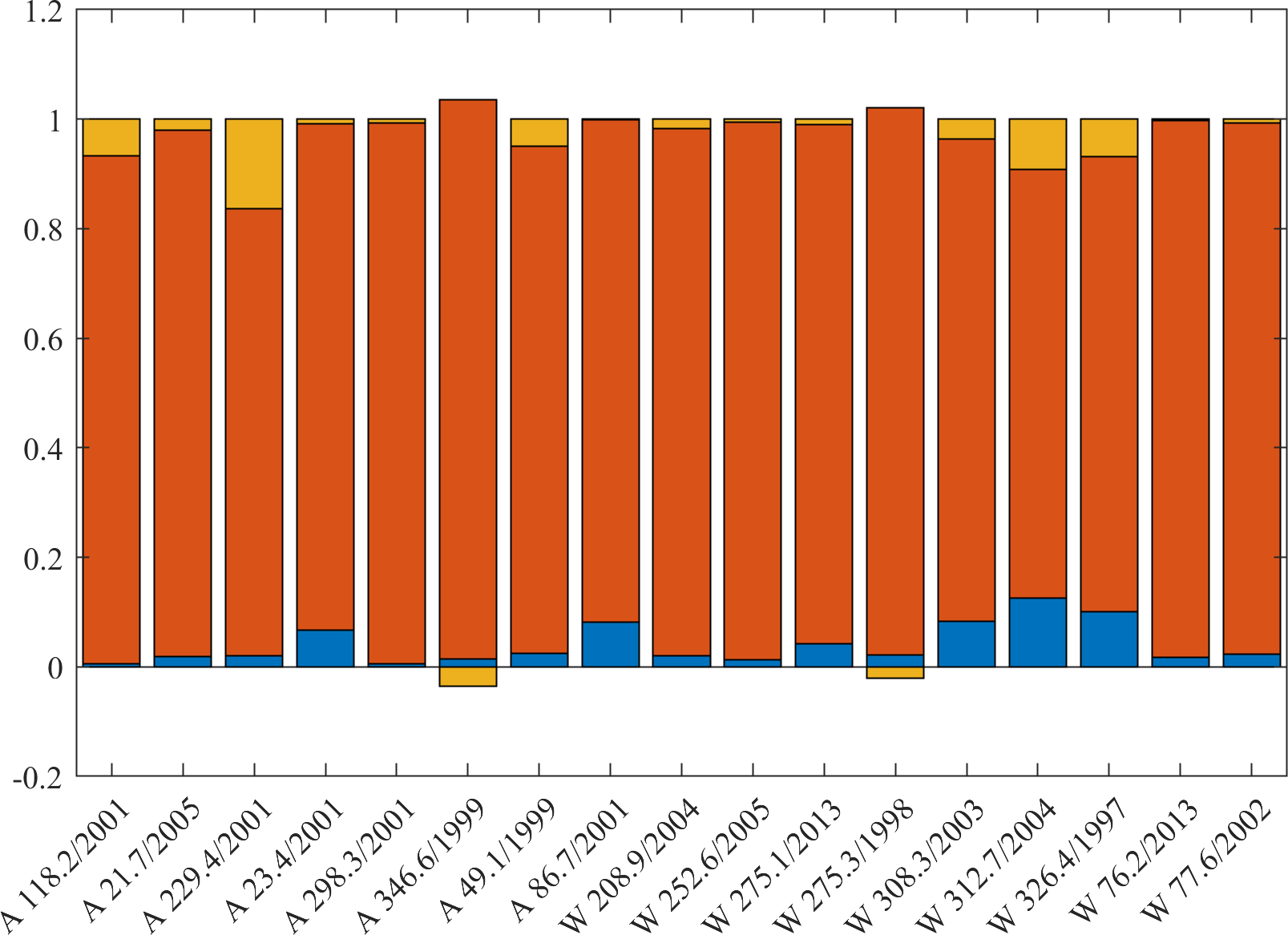}
         \caption{Upstream energy fractions.}
     \end{subfigure}
     \begin{subfigure}[b]{0.54\textwidth}
         \centering
         \includegraphics[width=\textwidth]{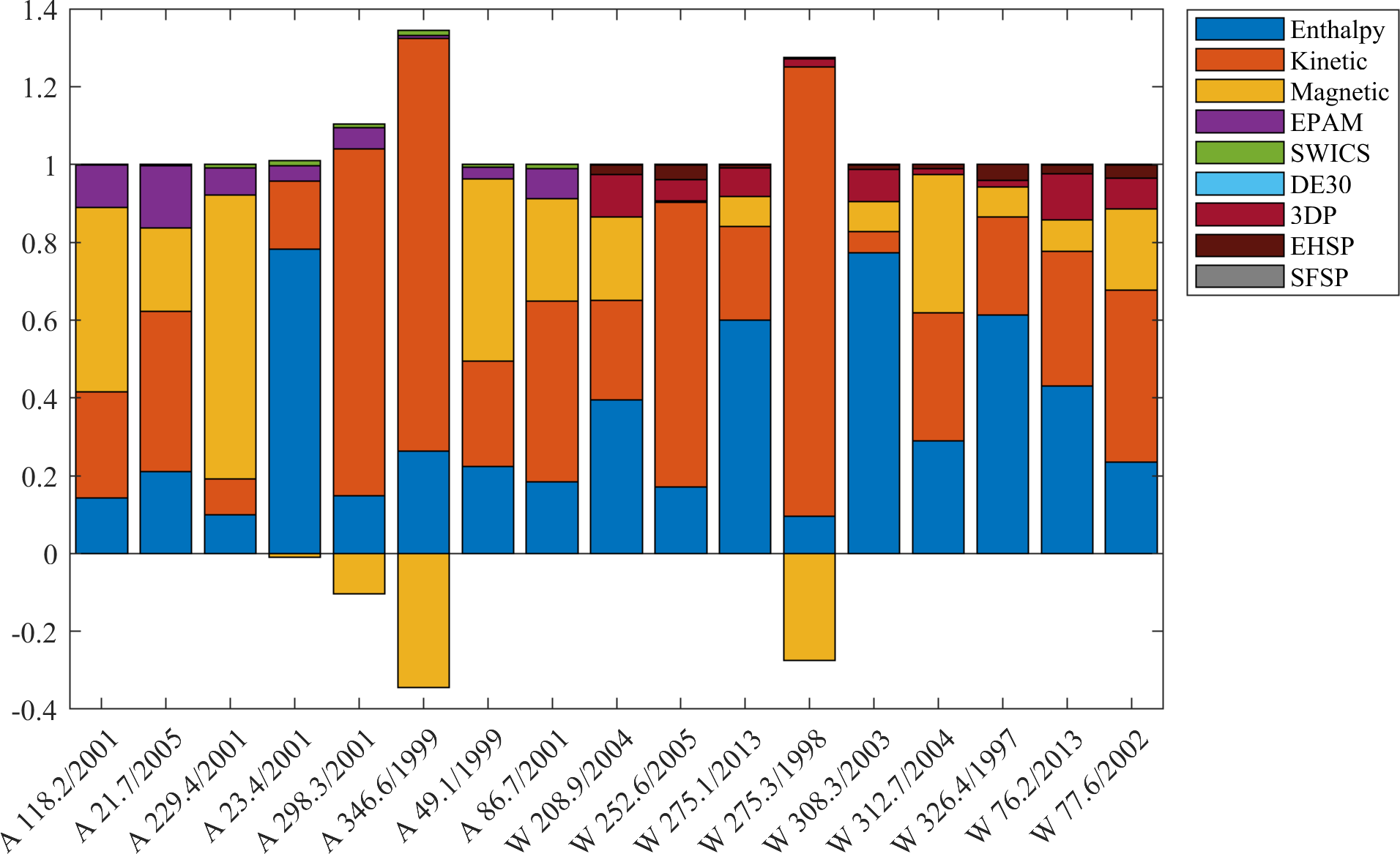}
         \caption{Downstream energy fractions.}
     \end{subfigure}
        \caption{The transfer of the shock kinetic energy (red) into both the plasma enthalpy (blue) and energetic particles (colors excluding yellow) is apparent in these plots. The total energy flux has been normalized to one. Negative magnetic fractions result from the $B_k(\textbf{v}\cdot\textbf{B})$ term in Equation 3. Some components are not visible due to their small fractions. \label{fig:energyfractions}}
\end{figure}

\section{Energetic Particle Fractions}\label{sec:EPF}

\begin{figure}
    \centering
\includegraphics[scale=0.55]{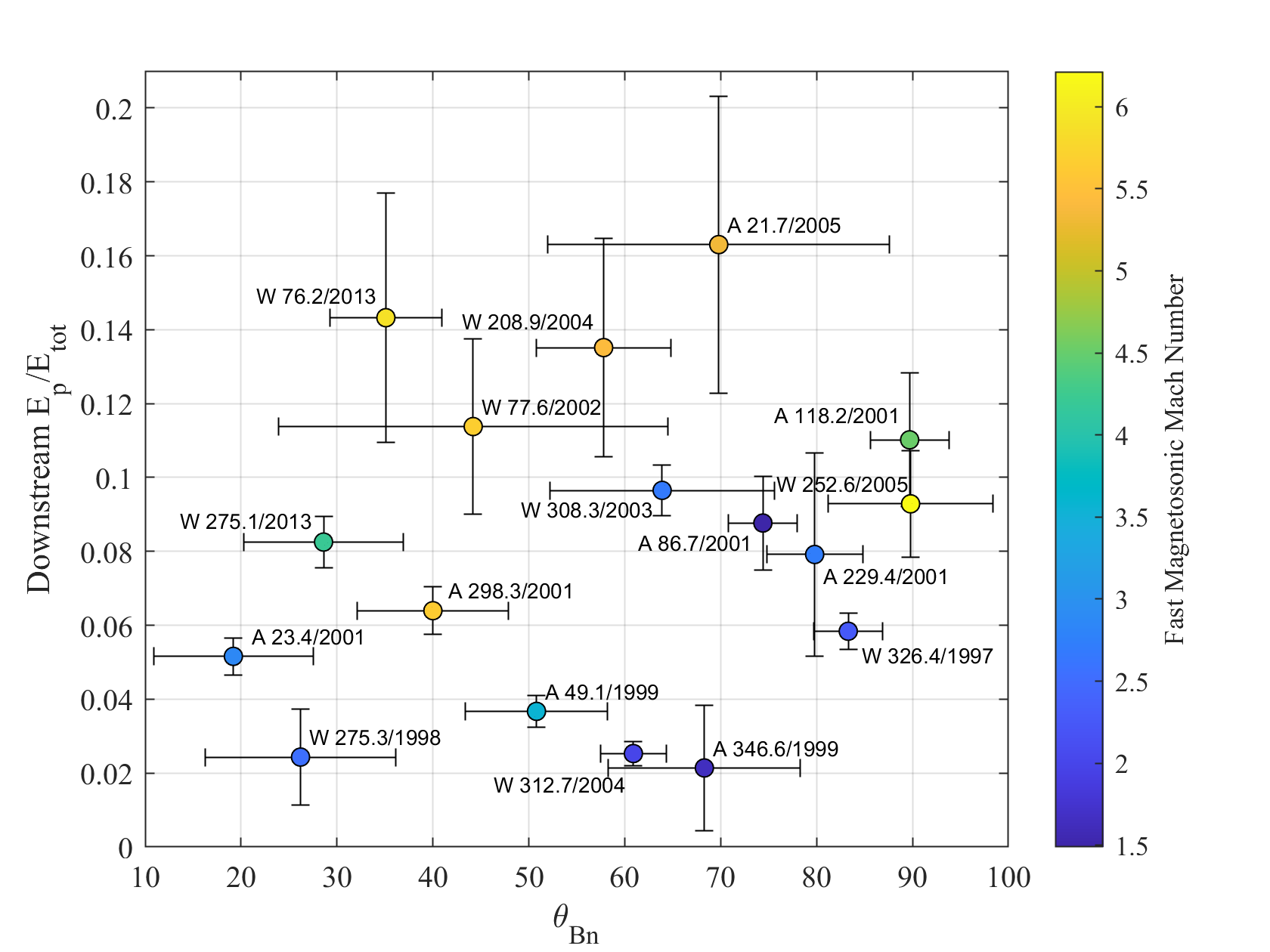}
    \caption{Ratio of the partial pressure of all energized particles (protons+electrons) to the total downstream energy density for all ACE and Wind events as a function of $\theta_{Bn}$. Each event is labelled with the DOY. Here the color scale designates the fast magnetosonic Mach number. \label{fig:particles}}
\end{figure}

The distinct contributions to the total energy fluxes for all events are shown in Figure \ref{fig:energyfractions}. Figure \ref{fig:particles} plots the total fraction of energy flux in downstream energetic particles. We find that the fractions range from 2.14\% to 16.3\%, with a mean of 8.15\% and median of 8.25\%. We conclude that the energetic particle fraction is $< 20\%$. Notably, the energy fraction is insensitive to both the fast magnetosonic Mach number and $\theta_{Bn}$, in stark contrast with some recent hybrid simulations (kinetic ions and fluid electrons), as summarized in the review \cite{Pohl.etal:2020}.

\begin{figure}
     \centering
     \begin{subfigure}[b]{0.45\textwidth}
         \centering
         \includegraphics[width=\textwidth]{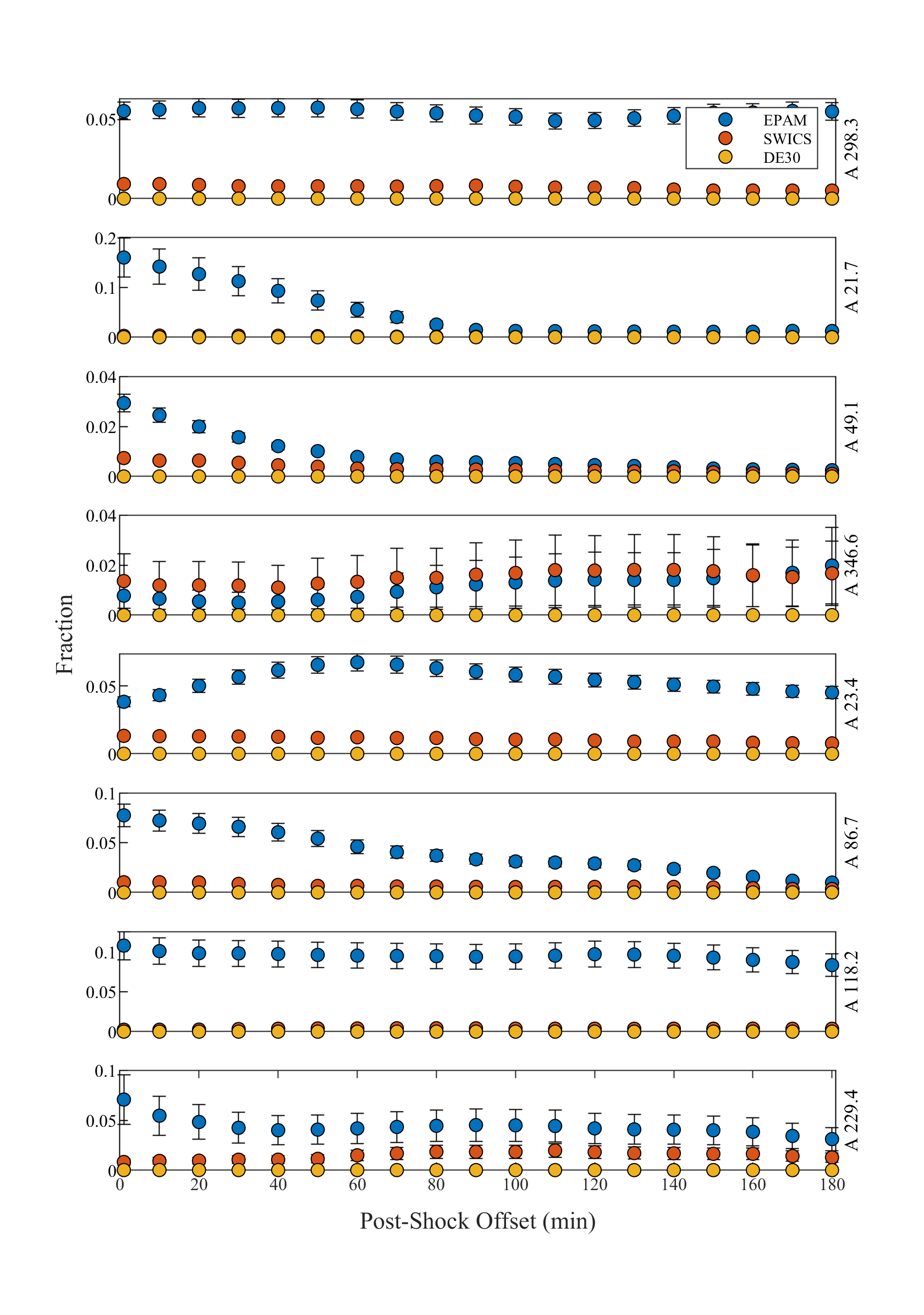}
     \caption{ACE shocks.}
     \end{subfigure}
     \begin{subfigure}[b]{0.45\textwidth}
         \centering
         \includegraphics[width=\textwidth]{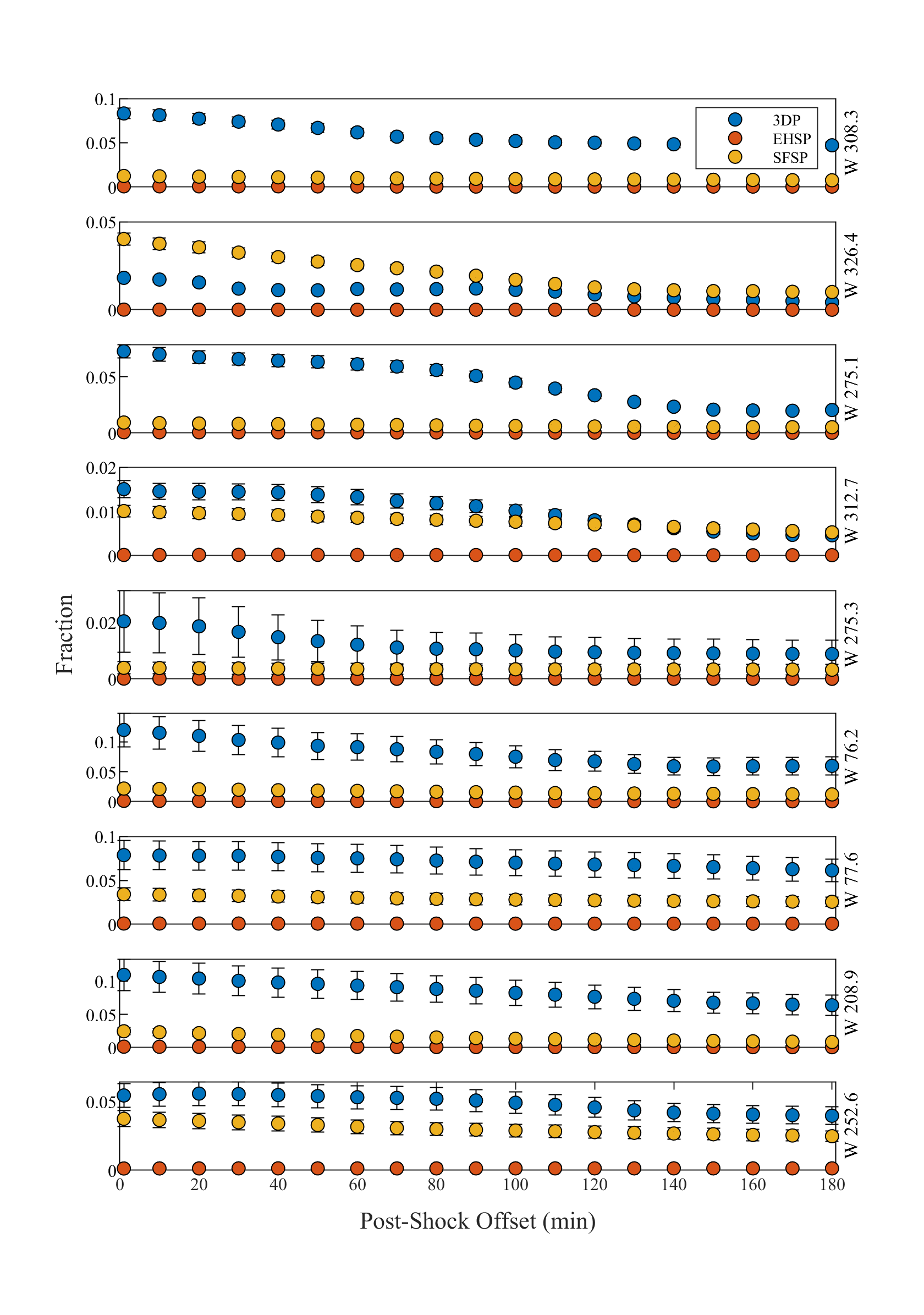}
         \caption{Wind shocks.}
     \end{subfigure}
        \caption{For each shock, with a label on the right vertical indicating the DOY ((a) for ACE and (b) for Wind events), the panels show how the fractional proton and electron energy flux contributions change with a 10-minute spacing of time-offsets in the one-hour particle flux average.\label{fig:timeoffset}}
\end{figure}

We tested the effect of shifting the one-hour energetic particle flux averages farther away from the shock time. Such an interval was offset in 10-minute steps between 10 and 180 minutes and included one minute past the shock time (see Fig.\ref{fig:timeoffset}). While one shock, ACE 118/2001, displays a consistent proton and electron energy flux fraction up to at least three hours downstream, the other events exhibit a variety of time-dependencies. For example, ACE 21/2005, 49/1999, 86/2016, and 229/2001, and Wind 326/1997, 275.1/2013, 275.3/1998, and 76.2/2013. exhibit a drop of about 50\% after one to two hours, while ACE 346/1999 increases with time. This latter event proceeds a much larger shock by about 0.2 days and was thus caught in the pre-shock exponential rise. ACE 23/2001 is maximum about 60 minutes after the shock, growing about 200\% from its initial value. Moreover, there is a short $\sim$15 minute spike in particle fluxes bridging the shock time, deviating from the exponential rise by a factor $\sim$2. This local enhancement is followed by an unusually slow rise in proton fluxes, meaning our one-minute offset, which was chosen as aforementioned to overlap maximally with CfA plasma quantities, precedes the peak.

\section{Discussion and Conclusion}

We have found that a significant fraction, up to 16\% but typically $\sim$8\%, of the energy flux downstream of an interplanetary shock is contributed by non-Maxwellian particle partial pressures. The consistency across $\theta_{Bn}$ does not support recent simulations, suggesting the role of large-scale pre-existing upstream magnetic turbulence in enhancing acceleration efficiency at quasi-perpendicular shocks \citep{Giacalone:2005:2,Fraschetti.Giacalone:15}. It has long been known that shocks with local $\theta_{Bn} > 45^\circ$ are efficient accelerators \cite{Jokipii:87}. The analysis presented herein shows direct evidence that the energy balance across the shock throughout the non-Maxwellian particle energy range requires their partial pressure to be accounted for at any $\theta_{Bn}$.

\acknowledgments

The material contained in this document is based upon work supported by a National Aeronautics and Space Administration (NASA) grant or cooperative agreement. Any opinions, findings, conclusions or recommendations expressed in this material are those of the author and do not necessarily reflect the views of NASA. We thank Dr. M. Stevens for confirming the Mach numbers and Dr. L. Wilson III for helping with Wind/PESA data. This work was supported through a NASA grant awarded to the Arizona/NASA Space Grant Consortium. FF was supported, in part, by NSF under grant 1850774, by NASA under Grants 80NSSC18K1213 and 80NSSC20K1283. JG was supported, in part, by NASA under Grants 80NSSC18K1213 and 80NSSC20K1283.

\end{document}